\documentclass[%
 reprint,
 amsmath,amssymb,
 aps,
]{revtex4-1}
\usepackage{graphicx}
\usepackage{dcolumn}
\usepackage{bm}
\usepackage{amsmath}
\usepackage{float}
\usepackage{multirow}
\begin{document}
\preprint{APS/123-QED}

\title{Three-Body Decay of $\Lambda_c^{*} (2595)$ and $\Lambda_c^{*} (2625)$  with Consideration of \\
$\Sigma_c(2455)\pi$ and $\Sigma_c^*(2520)\pi$ in Intermediate States}

\author{A. J. Arifi$^1$}
\author{H. Nagahiro$^{2,1}$}
\author{A. Hosaka$^{1,3}$}
\affiliation{ 
$^1$Research Center for Nuclear Physics (RCNP), Osaka University, Ibaraki, Osaka 567-0047, Japan\\
$^2$Department of Physics, Nara Women's University, Nara 630-8506, Japan\\
$^3$Advanced Science Research Center, Japan Atomic Energy Agency, Tokai, Ibaraki 319-1195, Japan}

\date{\today}

\begin{abstract}
Three-body decay of charmed baryons $\Lambda^*_c(2595)$ and $\Lambda^*_c(2625)$ into $\Lambda_c \pi \pi$ are studied with effective Lagrangians in which the coupling constants are extracted from the non-relativistic quark model. 
We take into account sequential processes going through $\Sigma_c (2455)$ and $\Sigma^*_c (2520)$ in intermediate states. 
The total decay widths are sensitive to the coupling of $\Lambda^*_c$ with $\Sigma_c \pi$ open channel and to $\Sigma^*_c \pi$ closed channel. 
We find that $\Lambda_c^*(2595)$ and $\Lambda_c^*(2625)$ with $\lambda$ mode assignment can explain nicely the experimental data.
We also show invariant mass distributions of $\Lambda_c^*(2595)$ and $\Lambda_c^*(2625)$ decays which are significantly different for various quark configurations.
\end{abstract}

\pacs{Valid PACS appear here}
\maketitle

\section{INTRODUCTION}

One of unique features of charmed baryons is that the two internal modes, the so called $\lambda$ and $\rho$ modes, split.
With one charm quark, the $\lambda$ mode corresponds to the motion of the two light quarks (diquark) relative to the charm quark, while the $\rho$ mode is the relative motion between the two light quarks with the charm quark regarded as a spectator~\cite{copley}. 
Generally, excitation energies of the $\lambda$ mode appear lower than those of the $\rho$ mode due to larger inertia mass.
This splitting has been known for long time as an isotope shift whose physical origin differs from the spin-spin hyperfine splitting.  
Relatively small excitation energies of low-lying charmed baryons seem to indicate the $\lambda$ mode dominance in those states.  
Yet the identification (or dominance) of those modes should be confirmed by other means in addition to the mass spectrum. 

To detect their different natures, it is useful to study various transition processes, in particular decays \cite{yan,cho,cho2,rosner,albertus,cheng,zhong}.
This is the issue that we would like to address in this paper. 
Recently, two of the present authors \cite{nagahiro} have studied two-body decays of charmed baryons.  
They have shown that the ratio of the $\Lambda^*_c \rightarrow \Sigma_c(2455)\pi$ and $\Lambda_c^* \rightarrow \Sigma_c^*(2520)\pi$ decays provides useful information on the structures of higher exited $\Lambda_c^*$ baryons.

\begin{figure}[t]
\centering
\includegraphics[scale=0.18]{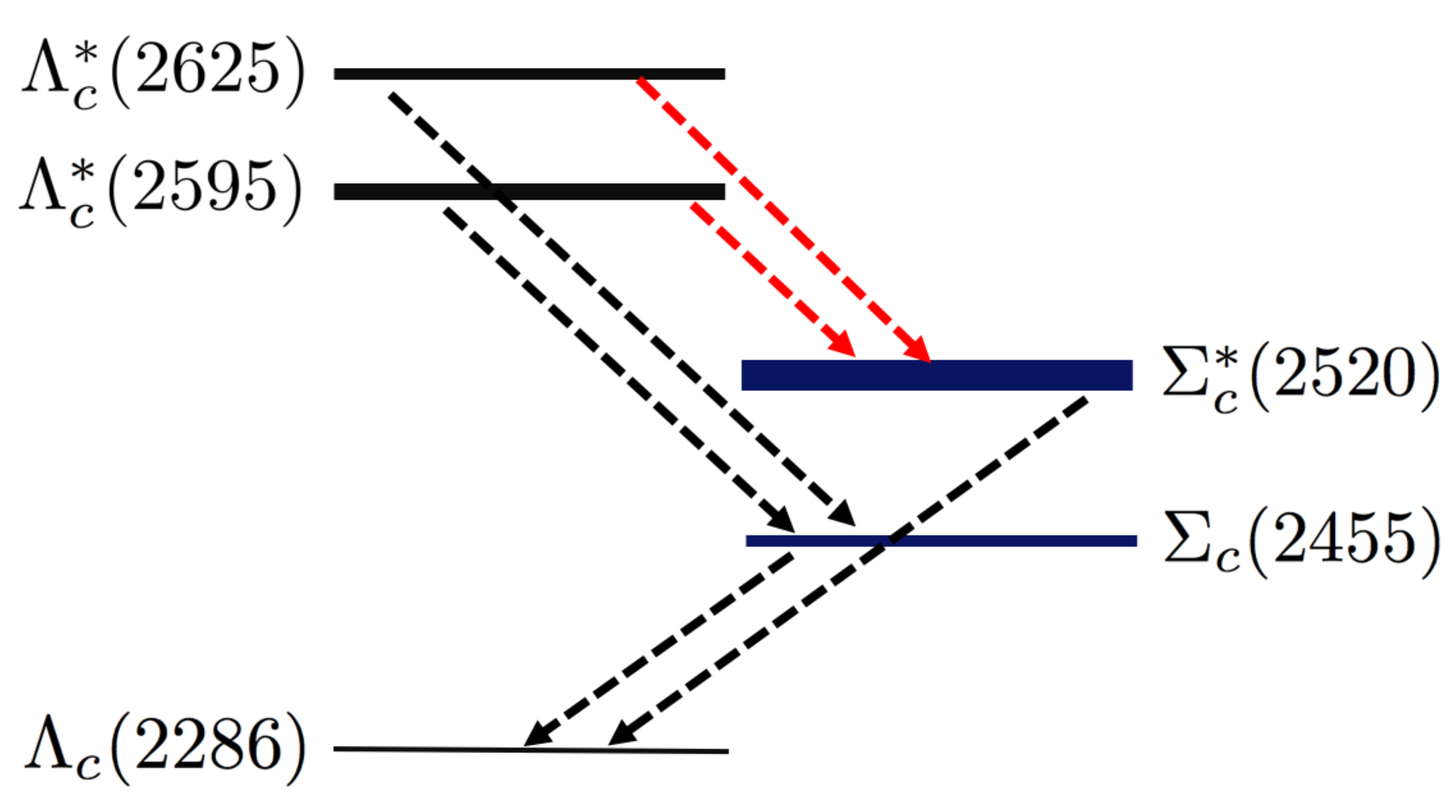}
\caption{Level structure of low lying charmed baryons and their strong decays through pion emission. Black arrows indicate kinematically allowed decays, while red arrows are for kinematically forbidden ones to the closed channel. }
\label{level}
\end{figure} 

For lower exited $\Lambda_c^*(2595)$ and $\Lambda_c^*(2625)$ states, the decay into $\Sigma_c(2455) \pi$ occur as a real process because the decaying channel is open, while the decay into $\Sigma_c^*(2520) \pi$ is not allowed because the channel is closed.  However, in the experimentally observed process where they measure $\Lambda_c \pi \pi$, the latter one may occur with $\Sigma_c^*(2520)$ as a virtual intermediate state. 
 According to PDG~\cite{pdg}, the decay of $\Lambda_c^{*}(2625)$ is dominated by the process quoted as ``$\Lambda_c \pi \pi $ 3-body" contribution. Here in this paper, we study the three-body decays going through  $\Sigma_c$(2455) and $\Sigma_c^*$(2520)  as an intermediate state, which we call sequential processes.   We will discuss that a large part of $\Lambda_c \pi \pi$ 3-body decay is explained by the sequential process through the closed $\Sigma_c^{*}(2520)$ channel for $\Lambda_c^*(2625)$ while its contribution to the decay of $\Lambda_c^{*}(2595)$ is small.  In this way, we can also extract the information on the closed channel.  
We also show the Dalitz plots and invariant mass distributions of $\Lambda_c^*(2595)$ and $\Lambda_c^*(2625)$ decays into $\Lambda_c\pi\pi$ for various quark configurations.
 This study is useful for further investigations of the structures of the charmed baryons.

This paper is organized as follows.  
In Sec. II, we formulate our method using the effective Lagrangians with various coupling constants determined by the quark model.  
In Sec. III, we discuss our numerical results compared to the experimental data. 
Finally, a summary is given in Sec. IV. We give detailed calculations for various amplitudes in Appendix A.

\section{FORMALISM}
\subsection{Effective Lagrangian}

\begin{figure}[t]
\includegraphics[width= 0.5\textwidth]{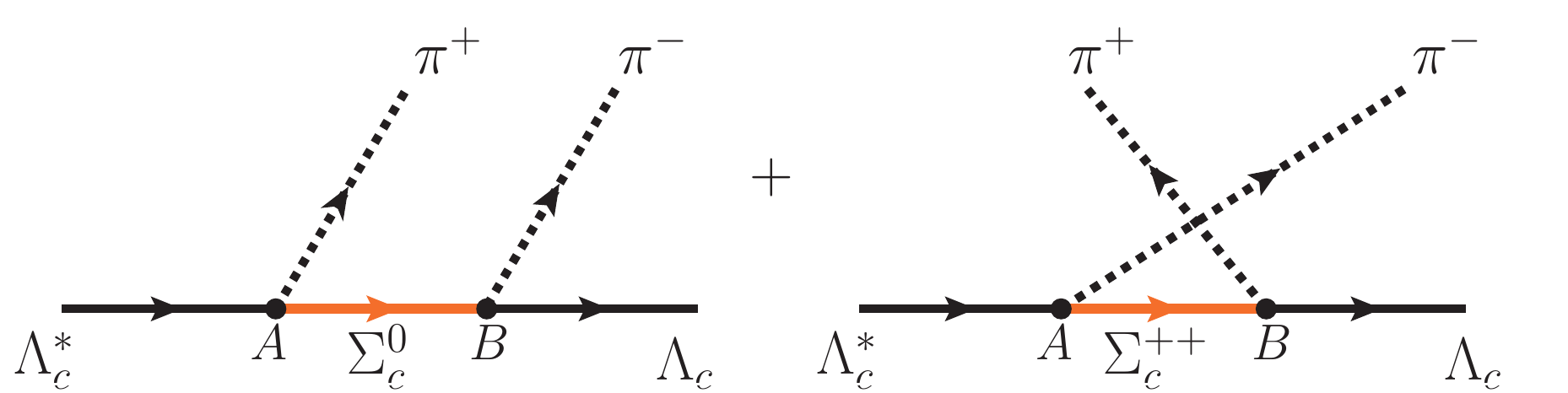}\\
\includegraphics[width= 0.5\textwidth]{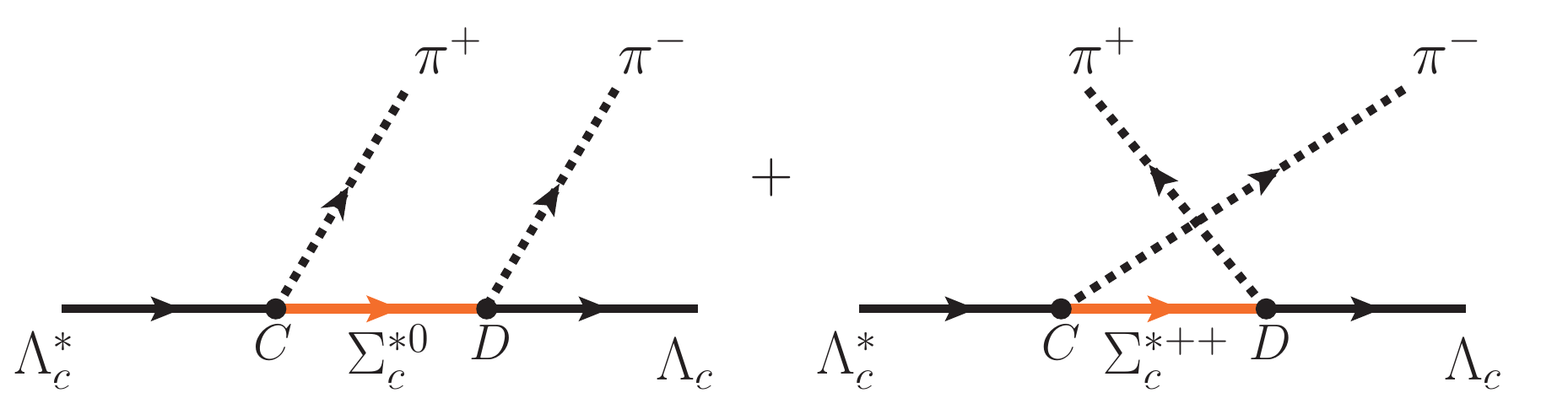} 
\caption{\label{diag34} All possible Feynman diagrams describing sequential decay of $\Lambda_c^{*} \rightarrow \Lambda_c \pi^+ \pi^-$. $\Lambda_c^{*}$  is either $\Lambda_c^{*} (2595)$ or $\Lambda_c ^{*}(2625)$. The diagrams consist of the process going through $\Sigma_c^*(2520)$ and $\Sigma_c(2455)$ and their different charged states.}
\label{diag}
\end{figure}

Let us discuss the two-pion emission decay amplitudes in the sequential process shown in Fig.~\ref{diag}. For this purpose, we introduce the effective Lagrangians describing the vertices of the diagrams. Our calculations are performed in the non-relativistic approximation which is considered to be good for the decays of charmed (heavy) baryons.  

For the case of $\Lambda_c^{*}(2595) \rightarrow \Lambda_c \pi^+ \pi^-$, the relevant Lagrangians are 
\begin{eqnarray}
\mathcal{L}_{A} &=& g_a \vec{\psi}^{\dagger}_{\Sigma_c} \psi_{\Lambda_c^*}\cdot \vec{\pi} + h.c, \label{1A}\\
\mathcal{L}_{B} &=& g_b \psi^{\dagger}_{\Lambda_c} \left(\vec{\sigma}\cdot \vec{\nabla}\right) \vec{\psi}_{\Sigma_c}\cdot \vec{\pi}+ h.c,\label{1B}\\
\mathcal{L}_{C} &=& g_c \vec{\psi}^{\dagger}_{\Sigma^*_c} \left(\vec{S}^{\dagger}\cdot \vec{\nabla} \vec{\sigma}\cdot \vec{\nabla} -\frac{1}{3} \vec{S}^{\dagger}\cdot \vec{\sigma} \vec{\nabla}^2\right) \psi_{\Lambda_c^*}\cdot \vec{\pi}+ h.c, \label{1C} \quad\\
\mathcal{L}_{D} &=& g_d {\psi}^{\dagger}_{\Lambda_c} \left(\vec{S}\cdot \vec{\nabla}\right) \vec{\psi}_{\Sigma^*_c} \cdot \vec{\pi}+ h.c, \label{1D}
\end{eqnarray}
where the derivatives act on the pion field, and the isovector structure is indicated explicitly for the pion and $\Sigma_c$ fields as $\vec{\pi}$ and $\vec{\psi}_{\Sigma_c}$. The structure of the Lagrangian also depends on the spin and parity of $\Lambda_c^*$ and hence also on the angular momentum of the out-going pion.  For instance, the vertex $A$ has $s$-wave structure, vertex $B$ has $p$-wave structure, vertex $C$ has $d$-wave structure, and so forth. In Eqs.~(\ref{1C}) and (\ref{1D}), spin transfer matrix $S_\mu$~\cite{ericson} is defined by the Clebsh-Gordan coefficients
\begin{equation}
\left< 3/2\ \alpha | S_\mu |1/2\ \beta \right> = \left(3/2\ \alpha\ 1\ \mu | 1/2\ \beta \right).
\end{equation}
where $\alpha$ and $\beta$ are the spin state of a particle with spin 3/2 and 1/2 respectively.

For $\Lambda_c^{*}(2625) \rightarrow \Lambda_c \pi^+ \pi^-$, the Lagrangian for each vertex is written as
\begin{eqnarray}
\mathcal{L}_{A} &=& f_a \vec{\psi}^{\dagger}_{\Sigma_c}  \left(\vec{\sigma}\cdot \vec{\nabla} \vec{S}\cdot \vec{\nabla} -\frac{1}{3} \vec{\sigma} \cdot \vec{S} \vec{\nabla}^2\right) \psi_{\Lambda_c^*} \cdot \vec{\pi} +h.c,\ \label{2A}\quad\\
\mathcal{L}_{B} &=& f_b \psi^{\dagger}_{\Lambda_c} \left(\vec{\sigma}\cdot \vec{\nabla}\right) \vec{\psi}_{\Sigma_c} \cdot \vec{\pi} + h.c, \label{2B}\\
\mathcal{L}_{C}^s &=& f_c \vec{\psi}^{\dagger}_{\Sigma^*_c} \psi_{\Lambda_c^*} \cdot \vec{\pi} + h.c, \label{2C}\\
\mathcal{L}_{C}^d &=& f'_c \vec{\psi}^{\dagger}_{\Sigma^*_c} \left(\vec{\Sigma}\cdot \vec{\nabla} \vec{\Sigma}\cdot \vec{\nabla} -\frac{1}{3} \vec{\Sigma} \cdot \vec{\Sigma} \vec{\nabla}^2\right)\psi_{\Lambda_c^*} \cdot \vec{\pi} + h.c, \label{2CC}\\
\mathcal{L}_{D} &=& f_d \psi^{\dagger}_{\Lambda_c} \left(\vec{S}\cdot \vec{\nabla}\right) \vec{\psi}_{\Sigma^*_c} \cdot \vec{\pi} + h.c. \label{2D}
\end{eqnarray}
where $\Sigma_\mu$ are
\begin{equation}
\left< 3/2\ \alpha | \Sigma_\mu |3/2\ \beta \right> = \left(3/2\ \alpha \ 1\ \mu | 3/2\ \beta \right).
\end{equation}
We note that there are two possible structures for the vertex $C$; $s$-wave and $d$-wave. Later, we will notice that the $s$-wave Lagrangian gives large contributions compared to $d$-wave.

The coupling constants $g$ and $f$ in the effective Lagrangians are extracted from the quark model by equating the amplitudes in the two models. In the quark model, the amplitudes of $Y_i \rightarrow Y_f \pi$ which correspond the vertices in Fig.~\ref{diag} are calculated by
\begin{eqnarray}
-i\mathcal{T}^{QM} &&(2\pi)^4\delta^{(4)}(p_f-p_i)=\nonumber\\
 &&\int {\rm d}^4x  \left< Y_f  \right(J_f, s' \left) \pi | i \mathcal{L}_{\pi q q}(x)| Y_i \right(J_i, s \left) \right>.\quad\quad
\end{eqnarray}
where $Y_{i(f)}$ denote the initial (final) charmed baryons, and the $\pi qq$ interaction in the quark model is given in the form of the pseudovector coupling.  It contains a quark axial coupling $g_A^q$ for the coupling strength.  This will be fixed to be one for our calculations.  The detailed calculation can be found in Ref.~\cite{nagahiro}. Likewise, the matrix elements derived from effective Lagrangians are
\begin{eqnarray}
-i\mathcal{T}^{EL}(2\pi)^4 &&\delta^{(4)}(p_f-p_i) = \nonumber\\
&& \int {\rm d}^4x \left< Y_f  \right(J_f, s' \left) \pi | i \mathcal{L}_{\alpha}(x)| Y_i \right(J_i, s \left) \right>,\quad\quad
\end{eqnarray}
where the symbol $\alpha$ stands for $A, B, C$ or $D$.

\subsection{Coupling constants for $\Lambda_c^{*}(2595)$}

The coupling constants in the effective Lagrangians~(\ref{1A})-(\ref{2D}) extracted from the non-relativistic quark model for $\Lambda_c^{*}(2595)$ with $\lambda$ mode are given by
\begin{eqnarray}
g_a &=& G 
\Bigg\{ \left( \frac{-1}{\sqrt{2}}\right) C_1 a_\lambda+ \left( \frac{q }{3\sqrt{2}} \right) C_2 \frac{  q_\lambda }{a_\lambda} \Bigg\},\\
g_b &=& \left( \frac{1}{\sqrt{3}}\right) iG C_2,\\
 g_c &=& \left(\frac{-1}{\sqrt{6}} \right) \frac{G C_2}{a_\lambda} \left(\frac{M}{2m+M}\right), \\
  g_d &=& -i G C_2.
\end{eqnarray}
where $M$ and $m$ are the masses of the heavy and light quarks and $a_\lambda$ is the range of the Gaussian wave function of $\lambda$ coordinate. We define the constants $G$ as
\begin{eqnarray}
G &=&\frac{g_A^q}{2f_\pi}
\end{eqnarray}
where $g^q_A$ is the quark axial vector coupling constant and $f_\pi=93$ MeV is the pion decay constant. For simplicity, we also define $C_1$ and $C_2$
\begin{eqnarray}
C_1(\omega_\pi, q) &=&\frac{\omega_\pi}{m} F(q),\\
C_2 (\omega_\pi, q) &=& \left[ 2 + \frac{\omega_\pi}{2m} \left(1- \frac{M}{2m+M}\right) \right] F(q),
\end{eqnarray}
where $F(q)$ is a Gaussian form factor
\begin{eqnarray}
F(q) = e^{-q^2_\lambda/4a^2_\lambda}e^{-q^2_\rho/4a^2_\rho}.
\end{eqnarray}
Furthermore, $\omega_\pi$ and $q$ are energy and momentum of emitting pion at corresponding vertices. The momentum transfer for the $\lambda$ and $\rho$ mode are given by
\begin{eqnarray}
q_\lambda &=& q \left(\frac{M}{2m+M}\right),\\
q_\rho &=& \frac{q}{2}.
\end{eqnarray}

For the $\rho$ mode assignment with $j=1$, the coupling constants become
\begin{eqnarray}
g_a &=& G 
\Bigg\{ 2C_1 a_\rho+ \left( \frac{-q }{3} \right) C_2 \frac{  q_\rho }{a_\rho} \Bigg\},\\
 g_c &=& \left(\frac{-1}{2\sqrt{3}} \right) \frac{G C_2}{2a_\rho}, 
\end{eqnarray}
where $g_b$ and $g_d$ remain the same because they are not dependent on the initial state mode. 
Here, $j$ is the total spin of the two light quarks including their orbital angular momentum (brown muck spin). 
There is also another possibility being $\rho$ mode ($j=0$), however, this mode is forbidden by spin conservation of brown muck.

\subsection{Coupling constants for $\Lambda_c^{*} (2625)$}

In the case of $\Lambda_c^{*}(2625)$ with $\lambda$ mode, the coupling constants are given by
\begin{eqnarray}
f_a &=&  \left(\frac{-1}{\sqrt{6}} \right) \frac{G C_2}{a_\lambda} \left(\frac{M}{2m+M}\right),\\
f_c &=&  G 
\Bigg\{ \left( \frac{-1}{\sqrt{2}}\right) C_1 a_\lambda+ \left( \frac{q }{3\sqrt{2}} \right) C_2 \frac{  q_\lambda }{a_\lambda} \Bigg\},\\
f'_c &=& -\frac{1}{8} \left(\frac{\sqrt{2}}{3} \right) \frac{G C_2}{a_\lambda}\left(\frac{M}{2m+M}\right).
\end{eqnarray}
where $f_c$ and $f'_c$ belong to the coupling constant at vertex $C$ with $s$-wave and $d$-wave structure respectively.

For the assignment with $\rho$ mode $(j=1)$, the coupling constants are expressed by
\begin{eqnarray}
f_a &=&  \left(\frac{-1}{2\sqrt{3}} \right) \frac{G C_2}{2a_\rho},\\
f_c &=&  G 
\Bigg\{2  C_1 a_\rho+ \left( \frac{-q }{3} \right) C_2 \frac{  q_\rho }{a_\rho} \Bigg\},\\
f'_c &=& -\frac{1}{8} \left(\frac{1}{3} \right) \frac{G C_2}{2 a_\rho}.
\end{eqnarray}
For $\rho$-mode $(j=2)$, the coupling constants are given by
\begin{eqnarray}
f_a &=&  \left(\frac{\sqrt{3}}{2\sqrt{5}} \right) \frac{G C_2}{2 a_\rho},\\
f_c &=&  0,\\
f'_c &=& -\frac{1}{8} \left(\frac{1}{\sqrt{5}} \right) \frac{G C_2}{2 a_\rho}.
\end{eqnarray}

\subsection{Model Parameters}
In the quark model of harmonic oscillator, there are three model parameters; $m$ the light quark mass, $M$ the heavy quark mass, and $k$ the spring constant \cite{nagahiro}. The quark masses are fixed to be
\begin{eqnarray}
m = 350\, {\rm MeV}, \hspace{1cm} M = 1500\, {\rm MeV}.
\end{eqnarray}
We also adjust the spring constant $k$ such that the range parameters of the Gaussian wave functions are fixed to be
\begin{eqnarray}
a_\lambda = 400\, {\rm MeV}, \hspace{1cm} a_\rho = 290\, {\rm MeV}.
\end{eqnarray}
which reproduce the quark distribution about 0.5 fm as corresponding to the nucleon core size.

\subsection{Amplitudes}
Let us first calculate the amplitude of $\Lambda_c^{*}(2595) \rightarrow \Lambda_c \pi^+ \pi^-$. The process is described by the diagrams in Fig.~\ref{diag}. The amplitude of the first diagram is expressed schematically by
\begin{eqnarray}
-i\mathcal{T} \left[\Sigma_c^0\right] = - i \frac{\mathcal{T}_{\Sigma_c^0\rightarrow \Lambda_c\pi^-}\mathcal{T}_{\Lambda_c^* \rightarrow \Sigma_c^0\pi^+} }{m_{23}-m_{\Sigma_c^0}+\frac{i}{2}\Gamma_{\Sigma_c^0}}, \quad \quad \label{amp1}
\end{eqnarray}
while the other charged state process in the second diagram is written by
\begin{eqnarray}
-i\mathcal{T} \left[\Sigma_c^{++}\right] &=& -i\frac{ \mathcal{T}_{\Sigma_c^{++}\rightarrow \Lambda_c\pi^+}\mathcal{T}_{\Lambda_c^* \rightarrow \Sigma_c^{++}\pi^-} }{m_{13}-m_{\Sigma_c^{++}}+\frac{i}{2}\Gamma_{\Sigma_c^{++}}}. \label{amp2}\quad\quad
\end{eqnarray}
In Eqs.~(\ref{amp1}) and (\ref{amp2}), $m_{23}$ and $m_{13}$ are the invariant masses of the subsystem of particle (2, 3) and (1, 3) respectively. They are
\begin{eqnarray}
m_{23}^2 &=& (P-p_1)^2 = (p_2 +p_3)^2,\\
m_{13}^2 &=& (P-p_2)^2 = (p_1 +p_3)^2,\\
m_{12}^2 &=& (P-p_3)^2 = (p_1 +p_2)^2,
\end{eqnarray}
where  $P$ is the energy-momentum of the initial baryon, and we have shown also $m_{12}$ for completeness. The particle numbers 1, 2, 3 are for $\pi^+$, $\pi^-$ and $\Lambda_c$. The third and forth diagrams are calculated similarly. Then, the total amplitude is expressed by
\begin{eqnarray}
-i\mathcal{T} &=& -i\mathcal{T} \left[\Sigma_c^0\right] -i\mathcal{T}\left[\Sigma_c^{++}\right] -i\mathcal{T}\left[\Sigma_c^{*0}\right]  \nonumber\\
& &-i\mathcal{T} \left[\Sigma_c^{*++}\right].
\end{eqnarray}
We give detailed calculations of each amplitudes in Appendix B. 

In deriving the squared amplitudes, there are some angular dependence in total amplitudes for which we have used the angle average approximation
\begin{eqnarray}
(\vec{p}_1 \cdot \vec{p}_2)^2 \rightarrow \frac{1}{3}  |\vec{p}_1|^2 |\vec{p}_2|^2.
\end{eqnarray}
This angular dependance $\cos^2 \theta$ ($\theta$ is the angle between the two pion momenta, $\vec{p}_1$ and $\vec{p}_2$) comes from the $d$-wave nature of the coupling to the $\Sigma_c^*(2520)$.  This may be used to confirm the contribution from  $\Sigma_c^*(2520)$ in the sequential process. 
Detailed study of angular correlations will be studied elsewhere.   
After some calculations, the spin summed (averaged for the initial state) amplitude becomes
\begin{eqnarray}
\frac{1}{(2J+1)}&&\sum_{s,s'} |-i\mathcal{T}|^2 = |G_{\Sigma^0_c}|^2 |\vec{p}_{2}|^2 + \frac{2}{9} |G_{\Sigma^{*0}_c}|^2  |\vec{p}_{1}|^4  |\vec{p}_{2}|^2  \nonumber\\
& &+ |G_{\Sigma^{++}_c}|^2 |\vec{p}_{1}|^2 + \frac{2}{9} |G_{\Sigma^{*++}_c}|^2  |\vec{p}_{2}|^4  |\vec{p}_{1}|^2 \quad \quad
\end{eqnarray}
where we have defined the quantity $G$, for instance,
\begin{eqnarray}
G_{\Sigma^0_c} = -i g_a g_b  \frac{\sqrt{2 m_{\Lambda_c^{*+}}} \sqrt{2 m_{\Lambda_c^+}}}{m_{23}-m_{\Sigma_c^0}+ \frac{i}{2} \Gamma_{\Sigma_c^0}}.\label{constants}
\end{eqnarray}
In Eq.~(\ref{constants}), $G_{\Sigma^0_c}$ contains information about the coupling constants and propagator for the corresponding diagram. 

In fact, there is another possible decay channel $\Lambda_c^{*}(2595) \rightarrow \Lambda_c \pi^0 \pi^0$. Different from the charged state process, the neutral pions assigned to be particle 1 and 2 are indistinguishable. Accordingly, we divide the amplitudes by the symmetric factor after we take into account all the numbered diagrams. Then, the total amplitude for a given decay channel can be written as
\begin{eqnarray}
-i\mathcal{T} &=& -i\mathcal{T} \left[\Sigma_c^+\right] -i\mathcal{T} \left[\Sigma_c^{*+}\right],
\end{eqnarray}
and the resulting squared amplitude is
\begin{eqnarray}
\frac{1}{(2J+1)}\sum_{s,s'}&&|-i\mathcal{T}|^2 = \nonumber\\
&& |G_{\Sigma^+_c}|^2 |\vec{p}_{2}|^2 + \frac{2}{9} |G_{\Sigma^{*+}_c}|^2  |\vec{p}_{1}|^4  |\vec{p}_{2}|^2. \quad\quad
\end{eqnarray}

Similarly, we can derive the amplitude of the $\Lambda_c^{*}(2625)$ decay. In this case, we have to include both $s$-wave and $d$-wave nature of the Lagrangian in vertex $C$, $\mathcal{L}_C^s$ and  $\mathcal{L}_C^d$. The squared amplitude of $\Lambda_c^{*}(2625) \rightarrow \Lambda_c \pi^+ \pi^-$ is then given by
\begin{eqnarray}
\frac{1}{(2J+1)}\sum_{s,s'}&&|-i\mathcal{T}|^2 = \frac{2}{3} |F_{\Sigma_c^0}|^2 |\vec{p}_{1}|^4 |\vec{p}_{2}|^2 \nonumber\\
&&+\frac{2}{3} |F_{\Sigma_c^{++}}|^2 |\vec{p}_{2}|^4 |\vec{p}_{1}|^2 + \frac{2}{3} |F_{\Sigma_c^{*++}}^s|^2 |\vec{p}_{1}|^2 \nonumber\\
&&+ \frac{32}{3} |F_{\Sigma_c^{*++}}^{d}|^2 |\vec{p}_{2}|^4 |\vec{p}_{1}|^2+ \frac{2}{3} |F_{\Sigma_c^{*0}}^s|^2 |\vec{p}_{2}|^2  \nonumber\\
& & + \frac{32}{3} |F_{\Sigma_c^{*0}}^{d}|^2 |\vec{p}_{1}|^4 |\vec{p}_{2}|^2
\end{eqnarray}
where the quantity $F$ are defined similarly to $G$. For instance, $F_{\Sigma_c^0}$ is denoted by
\begin{eqnarray}
F_{\Sigma_c^0} = -i f_a f_b  \frac{\sqrt{2 m_{\Lambda_c^{*+}}} \sqrt{2 m_{\Lambda_c^+}} }{m_{23}-m_{\Sigma_c^0}+\frac{i}{2}\Gamma_{\Sigma_c^0}}.
\end{eqnarray}
For $\Lambda_c^{*}(2625) \rightarrow \Lambda_c \pi^0 \pi^0$, the squared amplitude reads
\begin{eqnarray}
\frac{1}{(2J+1)}\sum_{s,s'} &&|-i\mathcal{T}|^2 = \frac{2}{3} |F_{\Sigma_c^{+}}|^2 |\vec{p}_{1}|^4 |\vec{p}_{2}|^2  \nonumber\\
& & + \frac{2}{3} |F_{\Sigma_c^{*+}}^{s}|^2 |\vec{p}_{2}|^2  + \frac{32}{3} |F_{\Sigma_c^{*+}}^{d}|^2 |\vec{p}_{1}|^4 |\vec{p}_{2}|^2. \quad\quad
\end{eqnarray}

\subsection{Three-Body Kinematics}

\begin{figure}[b]
\centering
\includegraphics[scale=0.6]{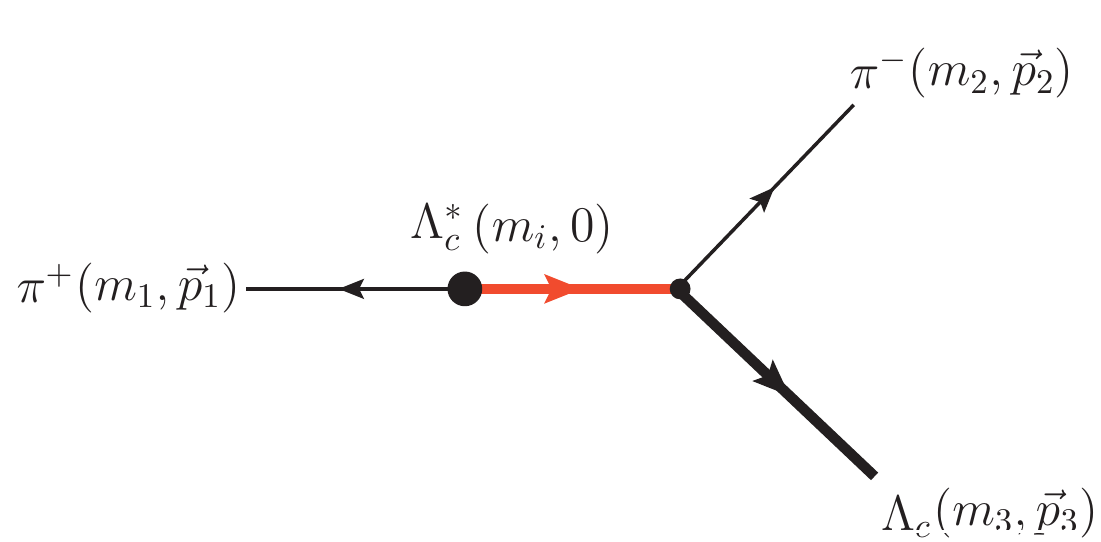}
\caption{Initial particle rest frame is considered in which the four momentum of the initial particle is $P=(m_i,\textbf{0})$. We define mass of $\pi^+, \pi^-, $ and $\Lambda_c$ as $m_1, m_2$ and $m_3$ respectively.}
\label{rest}
\end{figure}

\begin{figure}[H]
	\centering
	\includegraphics[scale=0.6]{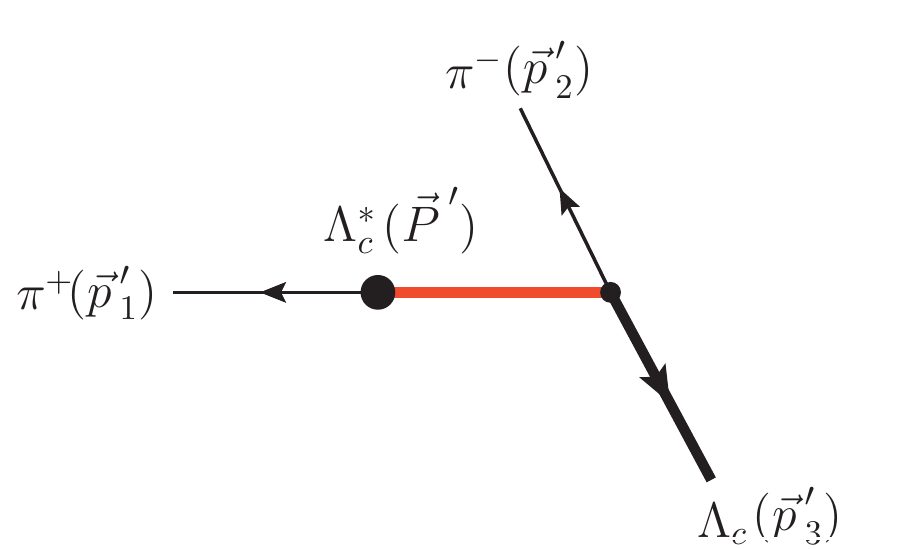}
	\caption{Resonance rest frame is defined as the rest frame of subsystem of particle 2 and 3.}
	\label{reso}
\end{figure}

The three-body decays are studied in the Dalitz plot in terms of the invariant masses $m_{12}$ and $m_{23}$~\cite{pdg}.  The actual momentum variables are defined in the rest frame of the initial $\Lambda_c^*$ as in Fig.~\ref{rest}, whereas various coupling constants are calculated in the rest frame of the intermediate $\Sigma_c$'s as in Fig.~\ref{reso}. 
The three-body decay widths are then given by 
\begin{eqnarray}
\Gamma &=& \frac{(2\pi)^4}{2m_i} \int \frac{1}{(2J+1)}\sum_{s',s}|-i\mathcal{T}|^2 {\rm d}\Phi_3(P;p_1,p_2,p_3),\nonumber\\
               &=& \frac{1}{(2\pi)^3}\frac{1}{32m_i^3} \int \frac{1}{(2J+1)}\sum_{s',s} |-i\mathcal{T}|^2 {\rm d}m^2_{13} {\rm d}m^2_{23}\quad\quad
\end{eqnarray}
where three-body phase space ${\rm d}\Phi_3$ in the first line depends on the 
initial energy square $s=m_i^2$, and the final state momenta, and is 
expressed by ${\rm d}m_{12}$ and ${\rm d}m_{23}$ in the second line.

\section{RESULTS AND DISCUSSIONS}

\subsection{Decay of $\Lambda_c^{*}(2595)$}

The $\Lambda_c^{*}(2595)$ baryon is the first excited state with spin and parity $J^P=1/2^-$ and its full width is $2.6\pm 0.6$ MeV. The $\Lambda_c^{*}(2595) \rightarrow \Lambda_c(2286)\pi\pi$ decay channel is the only possible strong decay~\cite{pdg,cdf}. Due to different excitation energies between the $\lambda$ and $\rho$ mode excitations, this state is expected to be dominated by the $\lambda$ mode~\cite{yoshida}. Here, we consider decays of both the $\lambda$ and $\rho$ modes to discuss the structure of $\Lambda_c^*(2595)$ from the view point of the decay property. $\Lambda_c^*(2595)$ can be constructed by the one $\lambda$ mode with $j=1$ and the two $\rho$ mode configurations of $j = 0,$ and 1. Detailed explanation of the configurations that we are using here can be found in our previous work \cite{nagahiro}.

\begin{table}[b]
\caption{\label{result12} Various contributions to the decay width of $\Lambda_c^{*}(2595) \rightarrow \Lambda_c\pi\pi$ in the sequential process with the $\lambda$ and $\rho$ mode assignments with different intermediate states (in unit of MeV). The right column shows partial decay widths into $\Sigma_c\pi$ and those into $\Lambda_c\pi\pi$ 3-body shown in PDG~\cite{pdg} }
\begin{ruledtabular}
\begin{tabular}{lccc|cc}
Intermediate & {$\lambda$-mode} 	& \multicolumn{2}{c}{$\rho$-mode} & &\multirow{2}{*}{Exp.}\\	
state &$j=1$&$j=0$&$j=1$&\\
\hline
$\Sigma_c^{++} \pi^-	$	&  0.237				& -	&	1.001		&		& 0.624 (24\%)\\
$\Sigma_c^0 \pi^+$		&  0.182 				& -	&	0.770		& 		&0.624 (24\%)\\
$\Sigma_c^{+} \pi^0$	&  1.629 				& -	&	6.896		& 		&- \\
					& 					&  	&				& 3-body &  0.468 (18\%)\\
$\Sigma_c^{*++} \pi^-$ 	&  1 $\times10^{-6}$  	& - 	& $6\times10^{-7}$	& 		&-\\
$\Sigma_c^{*0} \pi^+$ 	&  1 $\times10^{-6}$  	& - 	& $7\times10^{-7}$	& 		&-\\
$\Sigma_c^{*+} \pi^0$ 	&  5 $\times10^{-6}$  	& - 	& $3\times10^{-6}$ 	& 		&-\\
\hline
$\Gamma_{\rm total}$	&  2.048	 	& -	&  8.667		& &$2.6 \pm 0.6$\\
\end{tabular}
\end{ruledtabular}
\end{table}

The comparison between experimental data and calculated decay widths is presented in Table \ref{result12} where contributions from various intermediate states and with different mode assignments are shown separately. The upper three lines are the decays into $\Lambda_c\pi\pi$ from the open channels while the lower three lines those from closed ones. If we look at the total decay width in the bottom line, we find that the $\lambda$ mode assignment gives a consistent result with the experimental data. For the $\rho$ mode $(j = 1)$, the total decay width turn out to be broader and overestimate the data significantly. In contrast, $\rho$ mode $(j=0)$ assignment is forbidden due to the spin conservation of the brown muck as already pointed out in Ref.~\cite{zhong, nagahiro}.

\begin{figure}[t]
\centering
\includegraphics[width=0.49\textwidth]{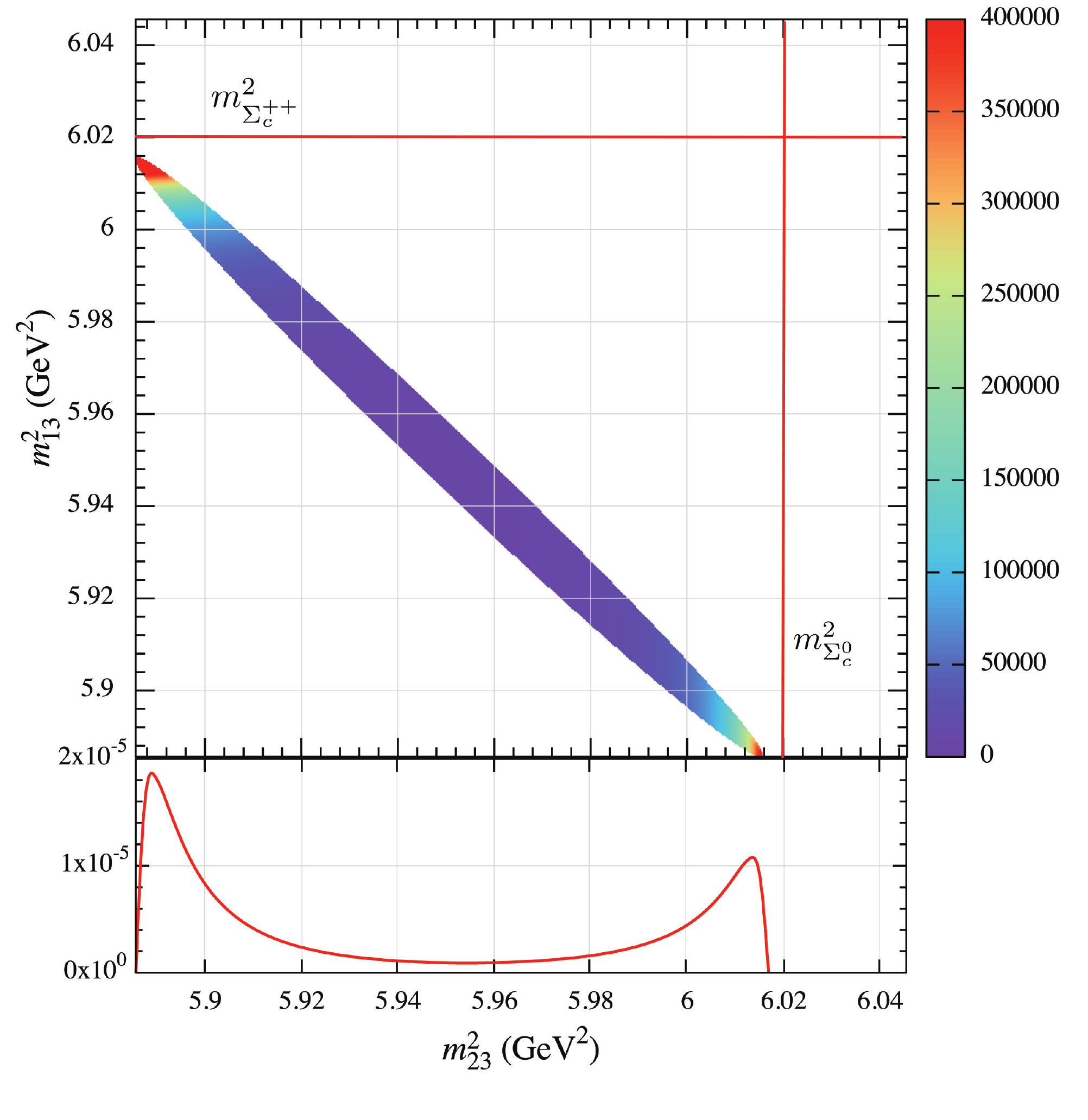}
\caption{ Dalitz plot and invariant mass distribution of  $\Lambda_c^{*}(2595) \rightarrow \Lambda_c \pi^+ \pi^-$ with $\Sigma_c^{(*)0}\pi^+$ and $\Sigma_c^{(*)++}\pi^-$ channels in intermediate state. $\Lambda_c^{*}(2595)$ is assumed to be $\lambda$ mode.}
\label{dalitz12}
\end{figure}

Isospin breaking effect can be seen clearly in both open ($\Sigma_c \pi$) and closed ($\Sigma_c^*\pi$) channels. In Table \ref{result12}, we can notice that the $\Sigma_c^+\pi^0$ channel contribution is larger than the other two charged channels. This is because the $\Sigma_c^{++}\pi^-$ and $\Sigma_c^{0}\pi^+$ channels are closed while the $\Sigma_c^+\pi^0$ channel is open, if we take the central values of the masses of $\Lambda_c^*(2595)$ and $\Sigma_c(2455)$.

\begin{figure}[H]
\centering
\includegraphics[width=0.48\textwidth]{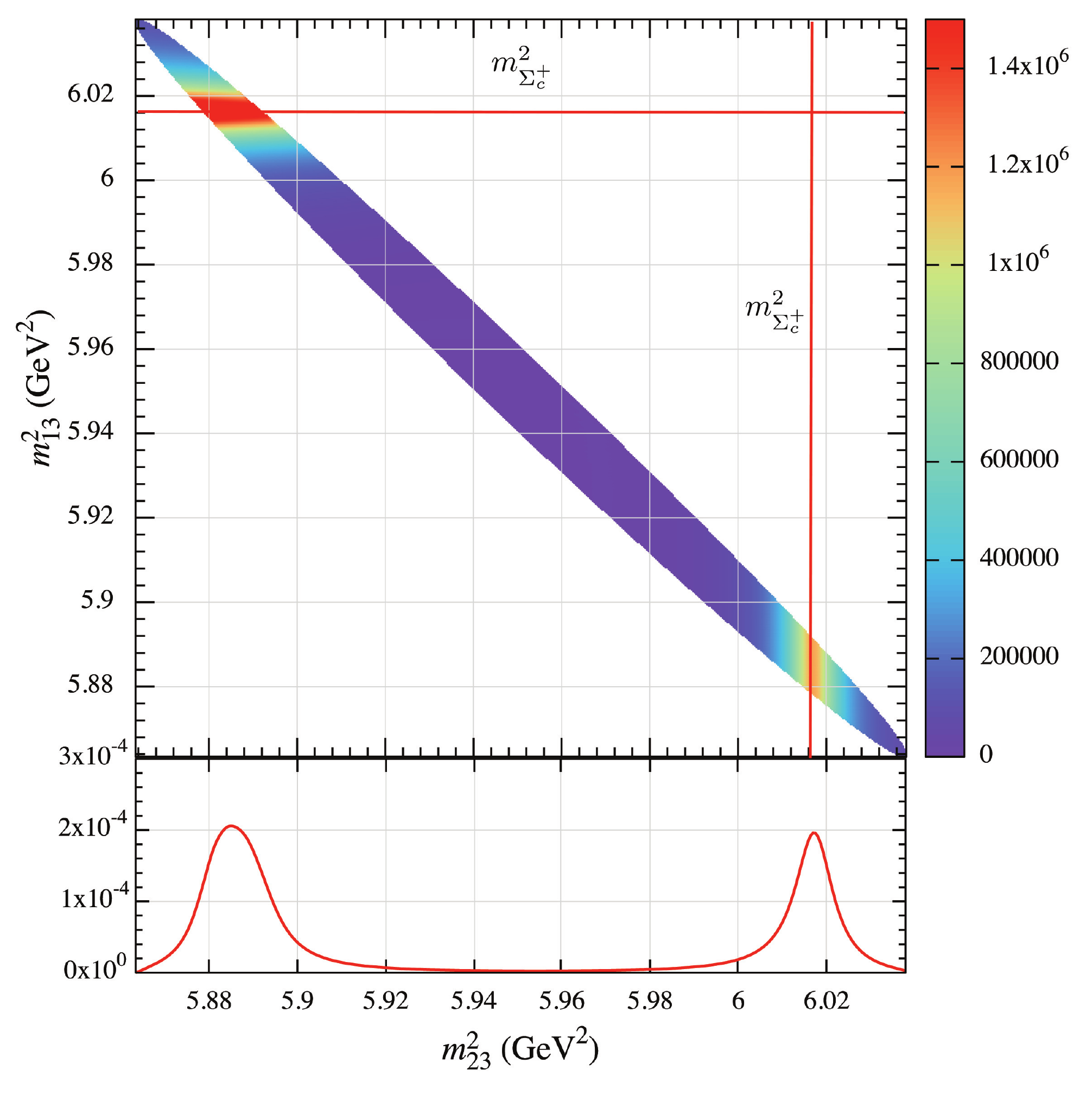}
\caption{ Dalitz plot and invariant mass distribution of  $\Lambda_c^{*}(2595) \rightarrow \Lambda_c \pi^0 \pi^0$ with $\Sigma_c^{(*)+}\pi^0$ channels in intermediate state. $\Lambda_c^{*}(2595)$ is assumed to be $\lambda$ mode.}
\label{dalitz12b}
\end{figure}

To help understand our sequential decay calculations better, in Fig.~\ref{dalitz12} we show Dalitz plot and invariant mass distributions for the squared amplitude for the charged pion decay mode as function of $m^2_{13} (= m^2_{\pi^+\Lambda_c})$ and $m^2_{23} (=m^2_{\pi^-\Lambda_c})$. 
There we see that the most events are concentrated in the boundary region of the maximum $m^2_{13}$ and $m^2_{23}$. 
These strengths come from the tail of the peak of $\Sigma_c(2455)$ which is located slightly outside the kinematically allowed region, as shown in Fig.~\ref{dalitz12}.
Similarly Fig.~\ref{dalitz12b} is for the neutral pion mode, where we see again the most events are near the boundaries but with the peak of $\Sigma_c(2455)$ is now inside the allowed region. 
This is due to isospin breaking effect, leading to the larger branching ratio for the neutral mode than the charged mode as shown in Table~\ref{result12} and the effect has also been discussed in Ref.~\cite{nagahiro}.

Let us turn to the discussion on the $\Sigma_c^*\pi$ contribution which is from the tail of the $\Sigma_c^*$ resonance.
The small contribution from the $\Sigma^*_c$ is expected not only from the fact that it is closed but also from the $d$-wave nature of the $\pi \Lambda(1/2^-)\Sigma^*(3/2^+)$ coupling. 

\subsection{Decay of $\Lambda_c^{*}(2625)$}

The $\Lambda_c^{*}(2625)$ baryon is the excited state having $J^P=3/2^-$. Experimentally, only the upper limit is given as $\Gamma_{\rm{exp}} < 0.97 $ MeV~\cite{pdg}. Different from the case of $\Lambda_c^*(2595)$, we cannot distinguish whether $\Lambda_c^{*} (2625)$ is $\lambda$ or $\rho$ mode by only looking at the two-body process with available experimental data. As shown previously~\cite{nagahiro}, the two-body contributions are too small as compared to the experimental upper limit.  This is due to the $d$-wave nature of the two-body final state.  Indeed, we show here that the three-body processes give significant contributions. This can be used to distinguish the $\lambda$ and $\rho$ modes.

\begin{table}[b]
\caption{ Various contributions to the decay width of $\Lambda_c^{*}(2625)$ in the sequential process with the $\lambda$ and $\rho$ mode assignments with different intermediate states (in unit of MeV). The right column shows partial decay widths into $\Sigma_c\pi$ and those into $\Lambda_c\pi\pi$ 3-body shown in PDG~\cite{pdg}  }
\begin{ruledtabular}
\begin{tabular}{lccc|cc}
 Intermediate& {$\lambda$-mode}  &  \multicolumn{2}{c}{$\rho$-mode} & &\multirow{2}{*}{Exp.{\cite{pdg}}}  \\
 state & $j=1$ & $j=1$ & $j=2$ &\\
\hline
$\Sigma_c^{++} \pi^-	$		& 0.037 	& 0.018 	& 0.033	& &$<$0.05 ($<$5\%) \\
$\Sigma_c^0 \pi^+$			& 0.031 	& 0.016	& 0.030 	& &$<$0.05 ($<$5\%) \\
$\Sigma_c^{+} \pi^0$		& 0.053 	& 0.027 	& 0.049 	&  &- \\
		 				& 		& 		& 		& 3-body&(large)\\
$\Sigma_c^{*++} \pi^-$ 		& 0.044 	& 0.190 	& 0 	& &-\\
$\Sigma_c^{*0} \pi^+$ 		& 0.064 	& 0.285 	& 0 	& &-\\
$\Sigma_c^{*+} \pi^0$ 		& 0.071 	& 0.306 	& 0 	& &-\\
\hline
$\Gamma_{\text{total}}$		& 0.300	& 0.842	& 0.112 	& &$< 0.97$\\
R						& 0.61	& 0.93	& 0		& \\
\end{tabular}
\end{ruledtabular}
\label{result32}
\end{table}

We compare decay widths calculated from various intermediates states with different mode assignments in Table~\ref{result32}. Firstly, our results are consistent with two-body analysis in our previous work~\cite{nagahiro} by which we can not disentangle which mode is dominant for $\Lambda_c^{*}(2625)$. Now by looking at the results of the closed channel contribution as shown in the lower three lines, we can see that the decay width is sensitive to the coupling of $\Sigma_c^*$.

Concerning the total decay width, experimentally only the upper limit is known. Therefore, we can not exclude all the possibility since they are below the upper limit. However, we analyze further by considering the ratio of the decay width
\begin{eqnarray}
R= \frac{\Gamma(\Lambda_c^*\rightarrow\Lambda_c\pi^+\pi^-(\text{non-resonant}))} {\Gamma(\Lambda_c^*\rightarrow\Lambda_c\pi^+\pi^-(\text{total}))},
\end{eqnarray}
where the value is $R=0.54\pm 0.14$~\cite{albrecht}. This value seems consistent with the $\lambda$ mode assignment.

\begin{figure}[b]
\centering
\includegraphics[width=0.47\textwidth]{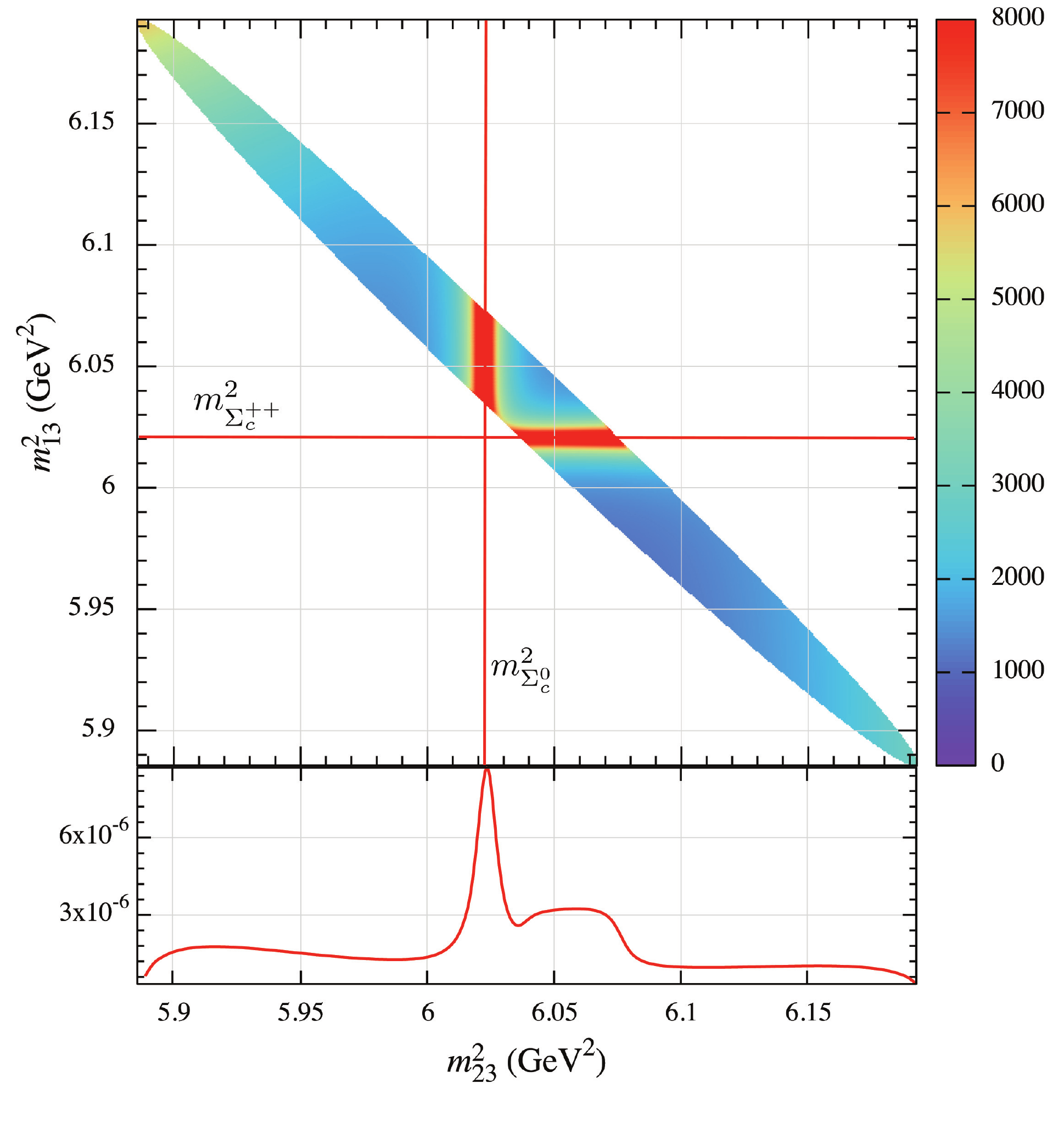}
\caption{\label{12lam3} Dalitz plot and invariant mass distribution of  $\Lambda_c^*(2625) (\lambda \rm{-mode})\rightarrow \Lambda_c \pi^+ \pi^-$}
\end{figure}

\begin{figure}[t]
\centering
\includegraphics[width=0.47\textwidth]{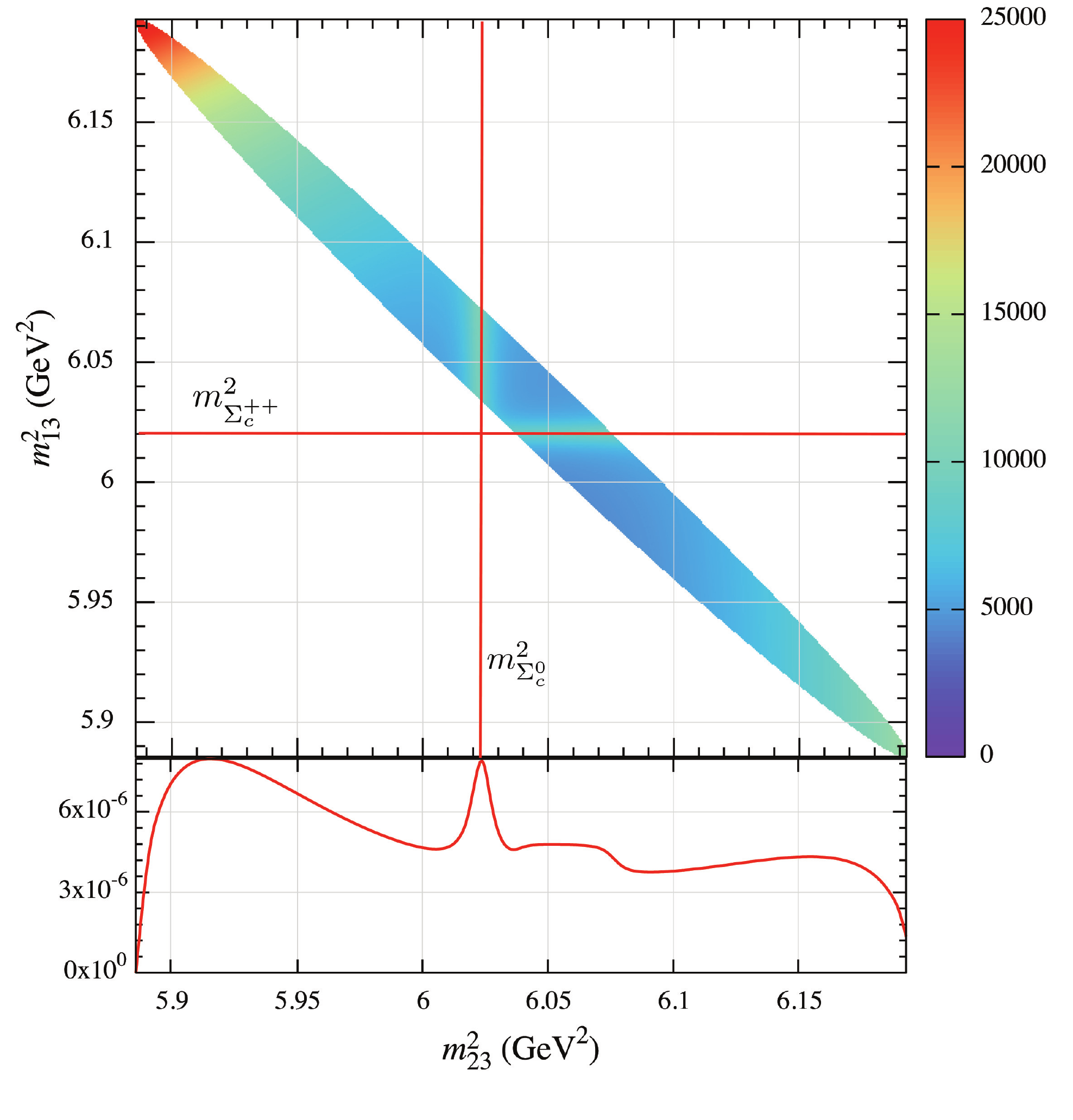}
\caption{\label{12lam4} Dalitz plot and invariant mass distribution of  $\Lambda_c^*(2625) (\rho \rm{-mode},$ $j=1)\rightarrow \Lambda_c \pi^+ \pi^-$}
\end{figure}

\begin{figure}[t]
\centering
\includegraphics[width=0.47\textwidth]{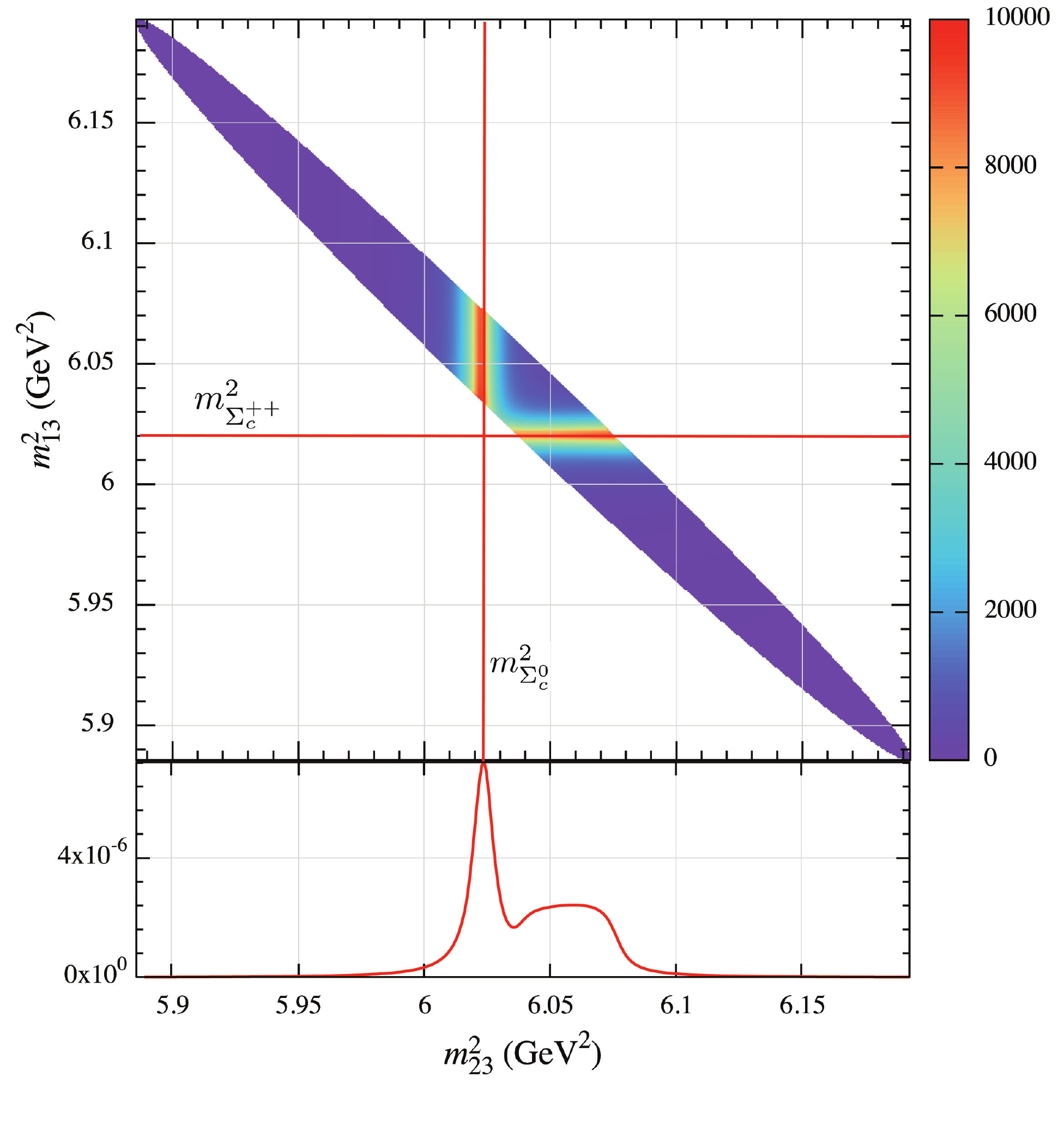}
\caption{\label{12lam5} Dalitz plot and invariant mass distribution of  $\Lambda_c^*(2625) (\rho \rm{-mode},$ $ j =2)\rightarrow \Lambda_c \pi^+ \pi^-$}
\end{figure}

In Fig.~\ref{12lam3}-\ref{12lam5}, we show Dalitz plots and invariant mass distributions for the decay of $\Lambda_c^*(2625)$ with different assignments. Fig.~\ref{12lam3} is for the $\lambda$ mode and shows that the most contributions are concentrated around the resonance $\Sigma_c(2455)$ region because the contribution of the closed channel $\Sigma_c^*(2520)$ is not large.

Figures~\ref{12lam4} and ~\ref{12lam5} are for the two $\rho$ modes, which show interesting features.  The $\rho$ mode with $j=1$ in Fig.~\ref{12lam4}  has a large contribution from the closed $\Sigma_c^*(2520)$ channel, showing a large background strength over the allowed region with less prominent peak structure from the open $\Sigma_c(2455)$ channel.  
Contrary, $\rho$ with $j=2$  mode has zero contribution from the closed channel and therefore, shows only a peak structure around the open $\Sigma_c(2455)$ channel.  These differences are clear, which will be useful to further distinguish the nature of the $\Lambda_c^*(2625)$ resonance.  


\section{SUMMARY}

Effective Lagrangian method has been used for the study of three-body decays of $\Lambda_c^{*}(2595)$ and $\Lambda_c^{*}(2625)$ in which the coupling constants are extracted from the quark model. We have considered the sequential decays through $\Sigma_c\pi$ and $\Sigma_c^*\pi$ in intermediate states. By comparing the theoretical predictions with the experimental data, we have extracted useful information about the excitation mode of those states.

By using currently available experimental data, we have argued that both $\Lambda_c^{*}(2595)$ and $\Lambda_c^{*}(2625)$ are most likely dominated by the $\lambda$ mode and all other possibilities of $\rho$ modes are unlikely. For $\Lambda_c^{*}(2595)$ with $\lambda$ mode, the two-body decay width is consistent with the data. In contrast, $\rho$ mode assignments overestimate significantly the decay width. In the case of $\Lambda^*_c(2625)$, the ratio of the $\Lambda^*_c \to \pi^+ \pi^-$ (non-resonant) and $\Lambda^*_c \to \pi^+ \pi^-$(total) decays seems consistent with the data, but further information on the Dalitz plots and invariant mass distributions should be useful to distinguish its structure.
\begin{acknowledgments}
This work is supported by a scholarship from the Ministry of Education, Culture, Science and Technology of Japan for A. J. A. and also Grants-in-Aid for Scientific Research (Grants No. JP26400275(C) for H. N.), (Grants No. JP26400273(C) for A. H.).
\end{acknowledgments}

\appendix
\section{Detailed Calculation of Amplitudes}

The decay of $\Lambda_c^{*}(2595) \rightarrow \Lambda_c \pi^+ \pi^-$ is described in Fig.~\ref{diag}. The amplitude for each diagram is given by
\begin{eqnarray}
-i\mathcal{T} \left[\Sigma_c^0\right] &=& -i \frac{\mathcal{T}_{\Sigma_c^0\rightarrow \Lambda_c\pi^-}\mathcal{T}_{\Lambda_c^* \rightarrow \Sigma_c^0\pi^+} }{m_{23}-m_{\Sigma_c^0}+\frac{i}{2}\Gamma_{\Sigma_c^0}} \quad \quad\nonumber\\
 &=& G_{\Sigma^0_c} \chi^{\dagger}_{\Lambda_c} \left(\vec{\sigma}\cdot \vec{p}_2\right) \chi_{\Lambda_c^*}
\end{eqnarray}
where
\begin{eqnarray}
G_{\Sigma^0_c} &=& -i g_a g_b  \frac{\sqrt{2 m_{\Lambda_c^{*+}}} \sqrt{2 m_{\Lambda_c^+}}}{m_{23}-m_{\Sigma_c^0}+\frac{i}{2}\Gamma_{\Sigma_c^0}}
\end{eqnarray}
Amplitude of the second diagram is
\begin{eqnarray}
-i\mathcal{T}  \left[\Sigma_c^{++}\right] &=& -i \frac{
\mathcal{T}_{\Sigma_c^{++}\rightarrow \Lambda_c\pi^+}\mathcal{T}_{\Lambda_c^* \rightarrow \Sigma_c^{++}\pi^-}}{m_{13}-m_{\Sigma_c^{++}}+\frac{i}{2}\Gamma_{\Sigma_c^{++}}} \nonumber\\
 &=& G_{\Sigma^{++}_c} \chi^{\dagger}_{\Lambda_c} \left(\vec{\sigma}\cdot \vec{p}_1\right) \chi_{\Lambda_c^*}
\end{eqnarray}
where
\begin{eqnarray}
G_{\Sigma^{++}_c} &=& -i g_a g_b  \frac{\sqrt{2 m_{\Lambda_c^{*+}}} \sqrt{2 m_{\Lambda_c^+}} }{m_{13}-m_{\Sigma_c^{++}}+\frac{i}{2} \Gamma_{\Sigma_c^{++}}}
\end{eqnarray}
and for the third diagram
\begin{eqnarray}
-i\mathcal{T} \left[\Sigma_c^{*0}\right] &=& -i  \frac{
\mathcal{T}_{\Sigma_c^{*0}\rightarrow \Lambda_c\pi^-}\mathcal{T}_{\Lambda_c^* \rightarrow \Sigma_c^{*0}\pi^+}}{m_{23}-m_{\Sigma_c^{*0}}+\frac{i}{2}\Gamma_{\Sigma_c^{*0}}}\nonumber\\
&=& G_{\Sigma^{*0}_c}  \chi^{\dagger}_{\Lambda_c} \left(\vec{S}\cdot \vec{p}_{2}\right) \times\nonumber\\
& &\left(\vec{S}^{\dagger}\cdot \vec{p}_{1} \vec{\sigma}\cdot \vec{p}_1 -\frac{1}{3} \vec{S}^{\dagger}\cdot \vec{\sigma} |\vec{p}_1|^2\right) \chi_{\Lambda_c^*}\nonumber\\
&=& G_{\Sigma^{*0}_c} \chi^{\dagger}_{\Lambda_c} \bigg(\vec{p}_1\cdot \vec{p}_2 \vec{\sigma}\cdot \vec{p}_1- \frac{1}{3} \vec{\sigma}\cdot \vec{p}_2 |\vec{p}_1|^2\bigg) \chi_{\Lambda_c^*}\nonumber\\\label{spin}
\end{eqnarray}
where
\begin{eqnarray}
G_{\Sigma^{*0}_c} &=& -i g_c g_d  \frac{\sqrt{2 m_{\Lambda_c^{*+}}} \sqrt{2 m_{\Lambda_c^+}} }{m_{23}-m_{\Sigma_c^{*0}}+\frac{i}{2} \Gamma_{\Sigma_c^{*0}}}.
\end{eqnarray}
In Eq.~({\ref{spin}}), we have used the spin matrix products
\begin{eqnarray}
S_i S_j^\dagger = \delta_{ij} -\frac{1}{3} \sigma_i \sigma_j.
\end{eqnarray}
The last amplitude reads
\begin{eqnarray}
-i&\mathcal{T}&\left[\Sigma_c^{*++}\right]= -i  \frac{\mathcal{T}_{\Sigma_c^{*++}\rightarrow \Lambda_c\pi^+}\mathcal{T}_{\Lambda_c^* \rightarrow \Sigma_c^{*++}\pi^-}}
{m_{13}-m_{\Sigma_c^{*++}}+\frac{i}{2} \Gamma_{\Sigma_c^{*++}}}\nonumber\\
&=& G_{\Sigma^{*++}_c}  \chi^{\dagger}_{\Lambda_c} \bigg(\vec{p}_2\cdot \vec{p}_1 \vec{\sigma}\cdot \vec{p}_2 - \frac{1}{3} \vec{\sigma}\cdot \vec{p}_1 |\vec{p}_2|^2 \bigg) \chi_{\Lambda_c^*}\quad\quad
\end{eqnarray}
where
\begin{eqnarray}
G_{\Sigma^{*++}_c} &=& -i g_c g_d  \frac{\sqrt{2 m_{\Lambda_c^{*+}}} \sqrt{2 m_{\Lambda_c^+}} }{m_{13}-m_{\Sigma_c^{*++}}+\frac{i}{2} \Gamma_{\Sigma_c^{*++}}}.
\end{eqnarray}

The total amplitude of the process is given by adding all the amplitudes coherently,
\begin{eqnarray}
-i\mathcal{T} &=& -i\mathcal{T} \left[\Sigma_c^0\right] -i\mathcal{T}\left[\Sigma_c^{++}\right] -i\mathcal{T}\left[\Sigma_c^{*0}\right]  \nonumber\\
& &-i\mathcal{T} \left[\Sigma_c^{*++}\right].
\end{eqnarray}
Therefore, the squared amplitudes consist of 16 terms which can be categorized into 5 contributions; The contribution from $\Sigma_c^0$,
\begin{eqnarray}
\frac{1}{(2J+1)}\sum_{s,s'}|-i\mathcal{T}|^2 \left[\Sigma_c^0\right] =  |G_{\Sigma^0_c}|^2 |\vec{p}_{2}|^2,
\end{eqnarray}
the contribution from $\Sigma_c^{++}$,
\begin{eqnarray}
\frac{1}{(2J+1)}\sum_{s,s'} |-i\mathcal{T}|^2\left[\Sigma_c^{++}\right] = |G_{\Sigma^{++}_c}|^2 |\vec{p}_{1}|^2,\quad
\end{eqnarray}
the contribution from the $\Sigma_c^{*0}$,
\begin{eqnarray}
\frac{1}{(2J+1)}\sum_{s,s'}|-i\mathcal{T}|^2 \left[\Sigma_c^{*0}\right] = \frac{2}{9} |G_{\Sigma^{*0}_c}|^2  |\vec{p}_{1}|^4  |\vec{p}_{2}|^2,\quad\quad
\end{eqnarray}
the contribution from $\Sigma_c^{*++}$,
\begin{eqnarray}
\frac{1}{(2J+1)}\sum_{s,s'} |-i\mathcal{T}|^2 \left[\Sigma_c^{*++}\right] = \frac{2}{9} |G_{\Sigma^{*++}_c}|^2  |\vec{p}_{2}|^4  |\vec{p}_{1}|^2.\nonumber\\
\end{eqnarray}
In addition, there are cross terms corresponding to the interference effects. However, all of the them vanish when we perform the angular integration for the total decay width.

For another decay channel $\Lambda_c^{*}(2595) \rightarrow \Lambda_c \pi^0 \pi^0$, there are only two process involved and the respective amplitudes are
\begin{eqnarray}
-i\mathcal{T} \left[\Sigma_c^+\right] &=& -i  \frac{\mathcal{T}_{\Sigma_c^{+}\rightarrow \Lambda_c\pi^0}\mathcal{T}_{\Lambda_c^* \rightarrow \Sigma_c^{+}\pi^0} }{m_{23}-m_{\Sigma_c^+}+\frac{i}{2} \Gamma_{\Sigma_c^+}} \nonumber\\
& & \nonumber\\
&=& G_{\Sigma^+_c} \chi^{\dagger}_{\Lambda_c} \left(\vec{\sigma}\cdot \vec{p}_2\right) \chi_{\Lambda_c^*}
\end{eqnarray}
where
\begin{eqnarray}
G_{\Sigma^{+}_c} &=& -i g_a g_b  \frac{\sqrt{2 m_{\Lambda_c^{*+}}} \sqrt{2 m_{\Lambda_c^+}}}{m_{23}-m_{\Sigma_c^{+}}+\frac{i}{2} \Gamma_{\Sigma_c^{+}}},
\end{eqnarray}
and 
\begin{eqnarray}
-i\mathcal{T} \left[\Sigma_c^{*+}\right] &=& -i \frac{\mathcal{T}_{\Sigma_c^{*+}\rightarrow \Lambda_c\pi^0}\mathcal{T}_{\Lambda_c^* \rightarrow \Sigma_c^{*+}\pi^0} }{m_{23}-m_{\Sigma_c^{*+}}+\frac{i}{2} \Gamma_{\Sigma_c^{*+}}}\nonumber\\
&=& G_{\Sigma^{*+}_c} \chi^{\dagger}_{\Lambda_c} \bigg(\vec{p}_1\cdot \vec{p}_2 \vec{\sigma}\cdot \vec{p}_1- \frac{1}{3} \vec{\sigma}\cdot \vec{p}_2 |\vec{p}_1|^2\bigg) \chi_{\Lambda_c^*}\nonumber\\
\end{eqnarray}
where
\begin{eqnarray}
G_{\Sigma^{*+}_c} &=& -i g_c g_d  \frac{\sqrt{2 m_{\Lambda_c^{*+}}} \sqrt{2 m_{\Lambda_c^+}} }{m_{23}-m_{\Sigma_c^{*+}}+\frac{i}{2} \Gamma_{\Sigma_c^{*+}}}.
\end{eqnarray}

The total amplitude is
\begin{eqnarray}
-i\mathcal{T} &=& -i\mathcal{T} \left[\Sigma_c^+\right] -i\mathcal{T} \left[\Sigma_c^{*+}\right]
\end{eqnarray}
The squared amplitudes now consist of 4 terms but the cross terms vanish again when we perform the angular integration. Therefore, only two terms contribute in the process. The first contribution from $\Sigma_c^{+}$ is
\begin{eqnarray}
\frac{1}{(2J+1)}\sum_{s,s'}|-i\mathcal{T}|^2 \left[\Sigma_c^{+}\right] =  |G_{\Sigma^+_c}|^2 |\vec{p}_{2}|^2,
\end{eqnarray}
the second contributions from $\Sigma_c^{*+}$ is
\begin{eqnarray}
\frac{1}{(2J+1)}\sum_{s,s'}|-i\mathcal{T}|^2 \left[\Sigma_c^{*+}\right] = \frac{2}{9} |G_{\Sigma^{*+}_c}|^2  |\vec{p}_{1}|^4  |\vec{p}_{2}|^2 \quad\quad
\end{eqnarray}

For the higher state, $\Lambda_c^*(2625)$, we calculate the decay amplitudes with the similar manner but with different spin structure which are derived from the Lagrangian in Eqs.~({\ref{2A}})-({\ref{2D}}). 


\nocite{*}
\bibliography{apssamp}

\end{document}